# Front End Data Cleaning And Transformation In Standard Printed Form Using Neural Models

Raju Dara[1], Dr.Ch. Satyanarayana[2], Dr.A.Govardhan[3]

[1]Scholar, JNTU, Kakinada, India
[2]Professor, JNTU, Kakinada, India
[3]Professor, JNTU Hyderabad, India

*Abstract*

*Front end of data collection and loading into database manually may cause potential errors in data sets and a very time consuming process. Scanning of a data document in the form of an image and recognition of corresponding information in that image can be considered as a possible solution of this challenge. This paper presents an automated solution for the problem of data cleansing and recognition of user written data to transform into standard printed format with the help of artificial neural networks. Three different neural models namely direct, correlation based and hierarchical have been developed to handle this issue. In a very hostile input environment, the solution is developed to justify the proposed logic.*

## 1. Introduction

Performing data cleaning on images is relatively a new research field and this process is computationally expensive on very large data sets, therefore e is was almost impossible to perform using conventional techniques the modern computers allow to perform the data cleansing process in acceptable time on large amounts of data. The quality of a large real world data set depends on various related issues, but the source of data is extremely vital aspect. Data quality issues can seriously skew the result of knowledge discovery from mining and analysis as a result that the corporations can make erroneous decisions based on misleading results and machinery could be incorrectly calibrated leading to disastrous malfunction which may potentially cost billions. If the data issues get unaddressed as a consequence they can be propagated rapidly across data systems, taints in an exponential manner. The window of opportunity to clean data is often quite tiny because the original data might be unavailable later to reconstruct a cleaned version. Data entry and acquisition is inherently prone to errors. Much effort could be given to this front-end process with respect to reduction in entry error but the fact frequently remains that the errors in a large data set are reasonably common unless an organization takes extreme measures to avoid errors. The field errors rates are typically around 5% or more occasionally. The logical solution to this problem is to be attempted to cleanse the data diplomatically which explores the data set for possible issues and endeavor to correct the errors. Of course, performing this task manually("by hand") for any real world data is completely out of the question and consumes considerable amount of time of a human. To avoid this sort of discrepancy, now a day's most of the companies spend millions of dollars per annum to discover and eradicate data errors. A manual approach for data cleansing is also laborious, and itself prone to errors. Hence it is required to have efficient and robust tools which automate or greatly assist in the data cleaning process that may be only the practical and cost effective way to achieve a reasonable quality level in an existing data set. This may seem to be an obvious solution while a very little research has been directly aiming at methods to support such tools. Some related research addresses the issues of data quality and tools to assist in by hand data cleaning and/or relational data integrity analysis

## 2. Role, and requirement of data cleaning and transformation at front end

Manual interaction practically associated with stimulus environment where there is an interest to collect necessary information with respect to required objects. Human observation and representation will not ensure the error free environment in practical approaches as a result it triggers a chain of mistakes for further. In some cases, this error may be tolerable but there the number of applications is coupled with it, since even a small mistake can cost huge or exceptionally hazardous for example health care, banking sector, defense services etc. This dictates the requirement to minimize the errors as much as possible. Thus there should be an automated solution to recognize the data and its entry in database. In the fig-1, path 1 shows the possible means to carry the error up to final destination where as path 2 has a module of data cleaning and eventually transformation in result must be more accurate.

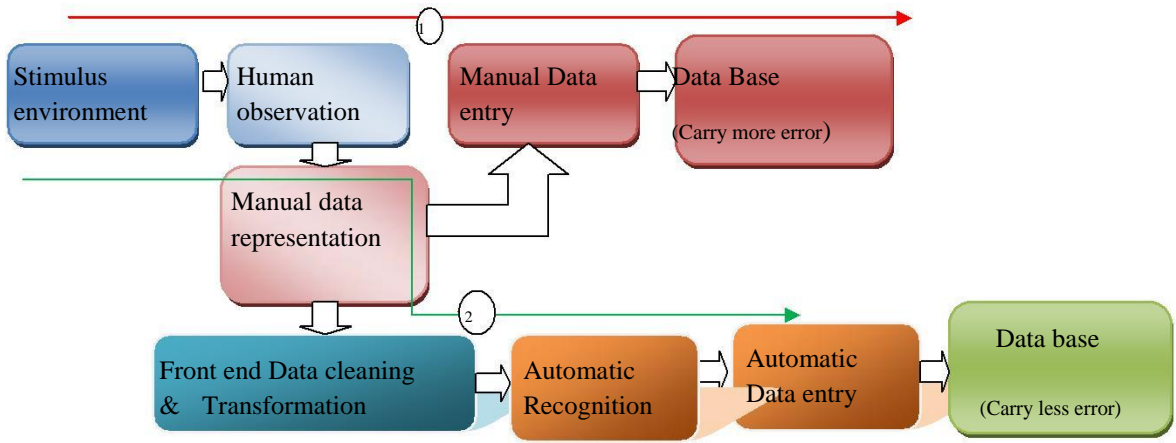

Figure-1. Front End Data Acquisition and Transferring process to data set

Objectives of this research is to develop a solution which could recognize and represent the handwritten information in standard printed form. So that action for system can be defined in written form or data can be stored in repository in standard form for further requirements.Complexity of problem is very obvious and challenging because of diversity in nature of human handwriting. Even the same person can be having a very less probability to write the exact pattern technically.Requirements and applications for such sort of solution are very important and numerous especially where data transformation is very precious or where there is a time constraint, e.g. in case of healthcare the formation about a patient who is in ICU and his related parametric information is being noted by hospital staff. Understanding of recorded information may be subjective and error may end with a disaster, the same mistake may cause huge cost in financial systems especially in banking and insurance sectors.

To handle this sort of problems, the requirement of ostensible intelligence and some standardized model of architecture is very much essential which could support the intelligence in better manner. At the same time it should supposed to take less amount of time to deliver the final result. With this constraint, the usage of artificial neural networks seems to be adaptive. Hence we have opted this technology as the solution platform. We have proposed three different models based on feedforward architecture to recognize the input character image pattern namely (1). Direct solution (2).Correlation based solution (3).Hierarchical solution. To make the input environment more hostile for developed solution, uncontrolled format of written characters have been chosen with manual selection so that the complexity of input could be maximized.

## 3.Artificial neural networks

A neural network is a parallel and distributed information processing structure consisting of processing elements (Neurons) interconnected with unidirectional signal channels called connections. Each neuron has a single output connection which branches into as many collateral connections as desired (each carries the same signal to the processing element output signal). The neuron's output signal can be of mathematical type. All of the processing that goes within each processing element must be completely local. Which means it must depend only on the current values of the input signal arriving at the processing element via impinging connections and the values stored in processing element's local memory. Neural networks has been started learning to earn their place among more conventional algorithmic approaches for solving many computational problems and are becoming important components of intelligence systems.Some of the properties which make neural networks appealing computational devices are:

i) Ability to learn and perform form example alone.
ii) Universality which makes it possible to use the same network models and learning algorithms across many diverse applications.
iii) The nonlinear nature of processing elements which gives the networks the ability to extract from important data regularities and salient features.
iv) Ability to represent knowledge naturally by means of network structure and state of neurons.
v) Ability to generalize from learned data to previously uncovered data.
vi) A dense interconnection network that provides a natural way in which neurons can provide context to each other.
vii) The graceful degradation of performance in the presence of damages to internal structure for some network architecture and learning problems.
viii) The simplicity of a network's neuron makes the good candidates for implementation in hardware
ix) The inherent massively parallel mode of analog computation which allows faster execution of specific tasks compared to traditional algorithmic implementation on digital computers.

International Journal on Computational Sciences & Applications (IJCSA) Vol.3, No.6, December 2013

## 4. Proposed Models

### 4.1 Direct Model

In this model a feedforward architecture having number of output neurons is equal to as many number of different input characters in training domain. Each output neuron corresponds to individual character which means all the characters get the learning over same weights and target defined according to the coded position of the each character in training domain. If any character in a particular position is an input to the neural network and its corresponding output neuron in output layer is expected to give the maximum (i.e., 1) where as all other output neurons to deliver minimum(i.e., equal to 0) value. Mean square error (MSE) is estimated to adjust the weights for further iteration.

### 4.2 Correlation based Model

A feed forward architecture with single output neuron in output layer is taken and each distinct input character obtains training individually.  In result for each character there is a dedicated set of trained weights. This can be considered as that set of trained weights to carry a particular type of relationship to deliver the corresponding output. Once training for all characters is over at the test time for a particular input character, each set of trained weights get the chance to interact with test input character and resulting output stored as degree of correlation between the test input character and that set of weights. Finally there are N numbers of degree of correlation. Decision of recognition for particular character is decided according to the position of maximal value of degree of correlation.

### 4.3 Hierarchical Model

There is a very good possibility of sharing some similarities among the possible domain of inputs. Hence based on the similarity, the characters form groups and each character holds a position in a systematic order. This model has two different parts namely (a) group recognizer (b) position recognizer.  In first part, training has given to learn the belonging group and in the next part, training has given to recognize the position of elements in the group. In result there is a set of weights for group recognizer and M number of set of weights for position recognizer. During examination for given test input,  the group is identified initially and later the corresponding character can be recognized with the help of corresponding position.

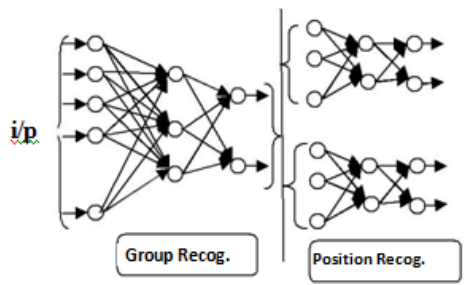

Figure-2. Hierarchical Structure

## 5. Experimental setup

### 5.1 Data Acquisition

Over a plane paper, English capital alphabets have been written number of times manually in row wise. The complete hand written paper has been scanned to convert it into an image format and the same is available as shown in the figure-3. With the hand cropping, individual character is isolated and preprocessing applied to extract the relevant pattern and its corresponding binary form. The complete process of pre-processing has been stated in next section. Size of cropped image is maintained as 25 X 20 pixels. Complete flow of process for pre-processing is shown in figure-4, where as the details of pattern extraction has been given in section 5.2. The position of character is defined as same as alphabetical order i.e., A-Z=1-26.

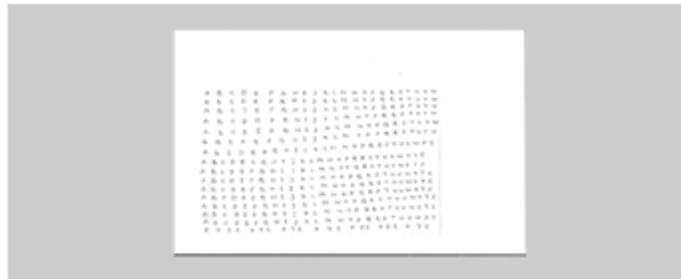

Figure-3. Image of User Written Data

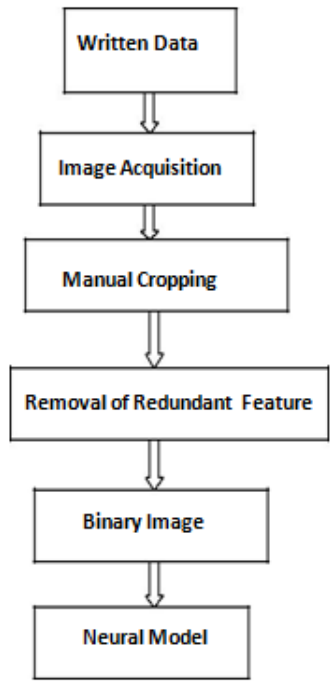

Figure-4. Flow of Process for Pre-processing

## 5.2 Pattern Extraction from Raw Cropped Image Block

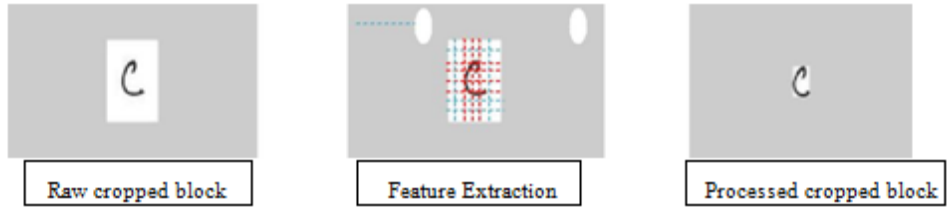

Figure-5. Flow Pattern Extraction

When a pattern is cropped by hand then there is a very good chance to have redundant information or very less information (poorly cropped), hence it is not possible to process the raw cropped image block directly. To overcome this issue an efficient mechanism has been developed which is based on concept of variation in standard deviation of row and column pixel values. Any row/column of block which carries any part of pattern then that row/column will definitely have higher value of standard deviation compared to row/column which does not contain any part of pattern as shown in fig.5. The row or column represented by 1 which does not contain any part of pattern whereas row or column represented by 2 contains the part of pattern hence std.dev(r/c 1)< std.dev(r/c 2). Once the redundant rows and columns are identified in cropped block, there are deleted from block until the required block size is not obtained as specified size say m×n. In the same scope if cropped block size is less than specified size then it is not accepted.

## 5.3 Architecture design

For all models fully interconnected three layered feedforward architecture has been developed. Number of input layers contain 500 neurons which are equal to the size of each input character image. Number of hidden layer neurons is made equal to 50 while output neurons are equal to 26, 1 and 11 for model1, model2 respectively, and group recognition part for model3. Unipolar sigmoid function is applied for active function in all active neurons. Target is set to 1 in corresponding position. Grouping in hierarchical model is decided according to similarity of patterns as shown in Figure-6 and as a result the number of neurons in second part of module3 is [2,4,3,4,2,3,1,3,1,2,1] as follows.

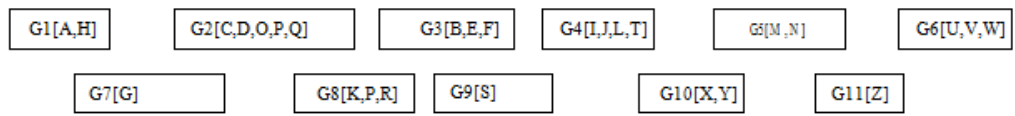

Figure-6.Grouping of characters for hierarchical model

## 5.4 Weight optimization with back propagation algorithm

Back propagation algorithm is a supervised learning algorithm which performs a gradient descent on a squared error energy surface to arrive at a minimum. The key to the use of this method on a multilayer perceptions is the calculation of error values for the hidden units by propagating the error backwards through the network. The local gradient for the $j^{th}$ unit, in the output layer is calculated as (assuming a logistic function for the sigmoid nonlinearity)

$$\delta_j = y_j(1- y_j)( d_j- y_j)$$

where $y_j$ is the output of unit j and $d_j$ is the desired response for the unit.
For a hidden layer, the local gradient for neuron j is calculated as

$$\delta_j = y_j(1- y_j) \sum_k \delta_{jk} w_{jk} \quad ------(1)$$

Where the summation k is taken over all the neurons in the next layer to which the neuron j serves as input. Once the local gradients are calculated each weight $w_{ji}$ is then altered according to the delta rule.

$$w_{ji}(t+1) = w_{ji}(t) + \eta \delta_j(t) y_i(t) \quad -------(2)$$

When η is a learning rate parameter and 't' is time. Frequent modification of equation (2) is used that incorporates a momentum term which helps to accelerate the learning process.

$$w_{ji}(t+1) = w_{ji}(t) + \eta \delta_j(t) y_i(t) + \alpha[w_{ji}(t) - w_{ji}(t-1)] --------(3)$$

Where α is a momentum term lying in the range $0 < \alpha < 1$

## 5.5 Parameter setting and performance

Terminating criteria: If SME < 0.001;
No of hidden models : 50
Learning rate : 0.2
Momentum constant : 0.1
Training data set size : First 10 rows of data equal to 260 characters;
Test data set size : last 5 rows of data equal to 130 characters;
Performance of all models has shown below for training and test cases.

**Model 1 performance:**

Learning process

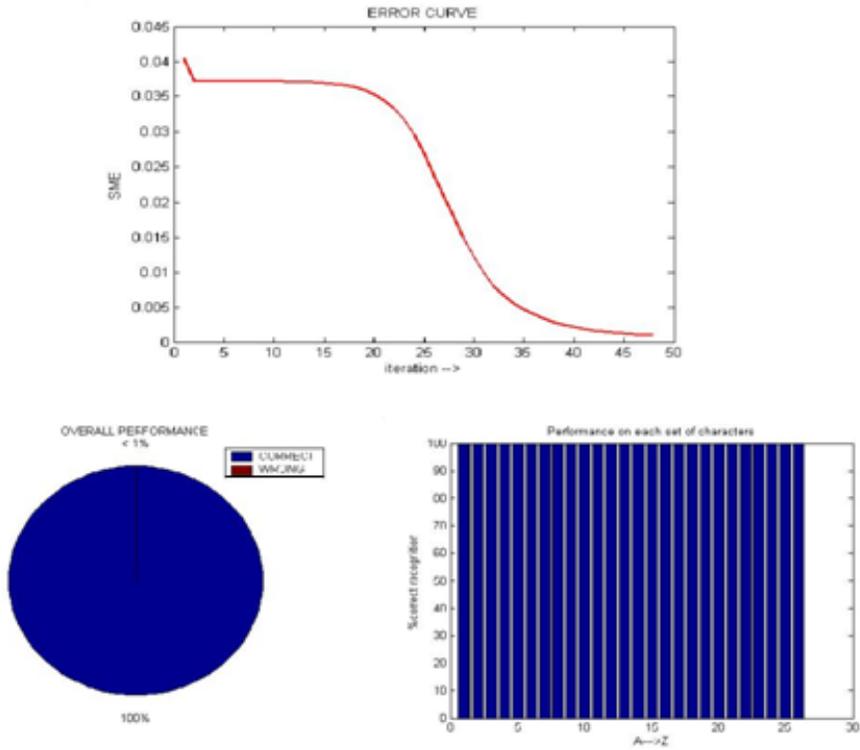

Figure-7 (a) Error reduction with learning process, (b) overall performance,(c) individual character performance.

Test performance

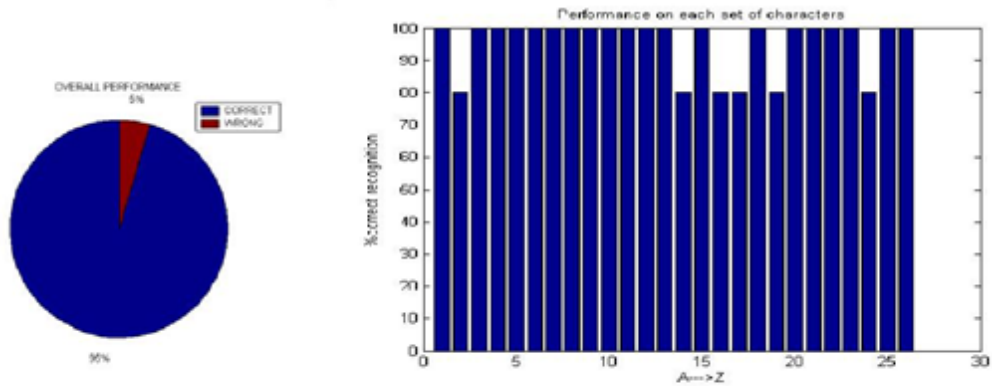

Figure-8 (a)overall performance,(b)individual character performance

**Model 2 performance :**

Learning process

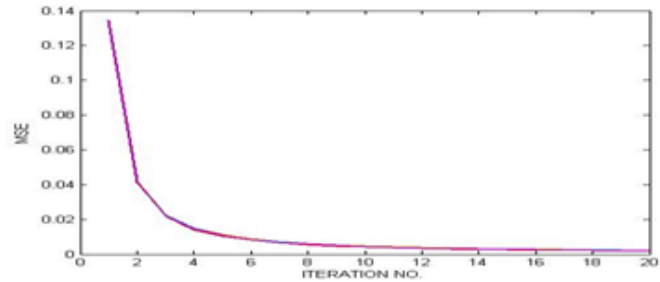

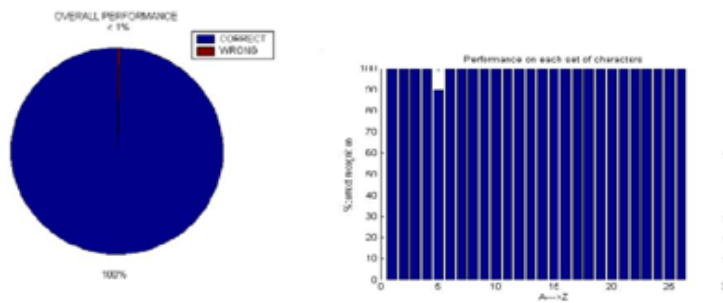

Figure-9. (a) Error reduction with learning process,(b)overall performance,(c)individual character performance

Test performance

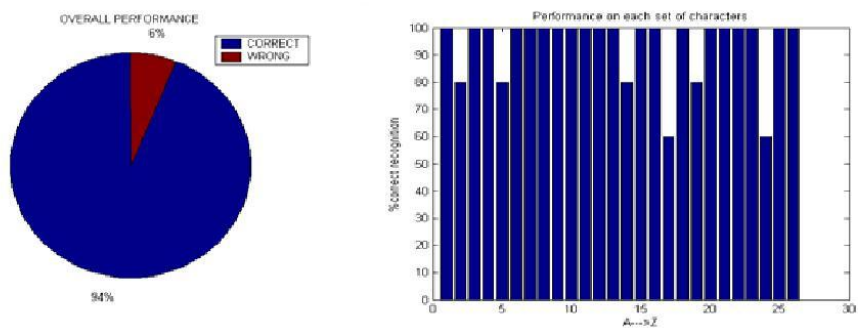

Figure-10.(a)overall performance,(b)individual character performance

**Model 3**

Learning performance

17

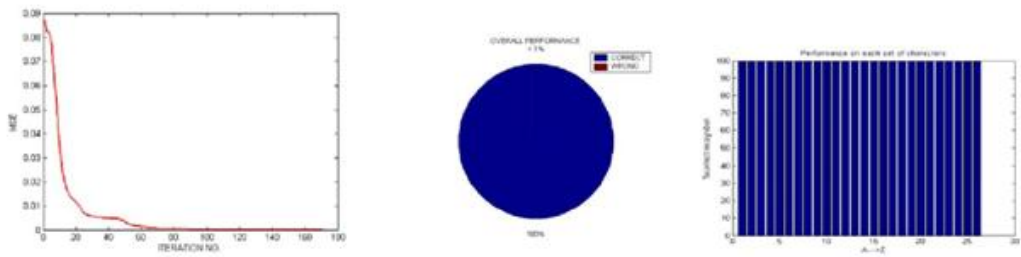

Figure-11. (a) Error reduction with learning process,(b)overall performance,(c)individual character performance

Test performance

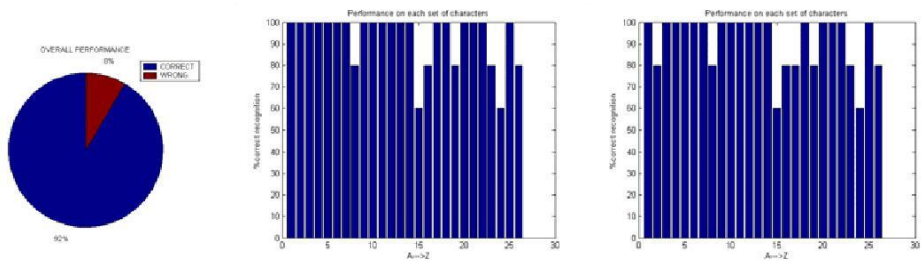

Figure-12. (a)overall performance, (b) performance in group recognition (c) performance in position recognition.

## An experiment with individual

Arbitrarily a character is cropped from test data set and all models are applied over that, performance shown by all models have given below

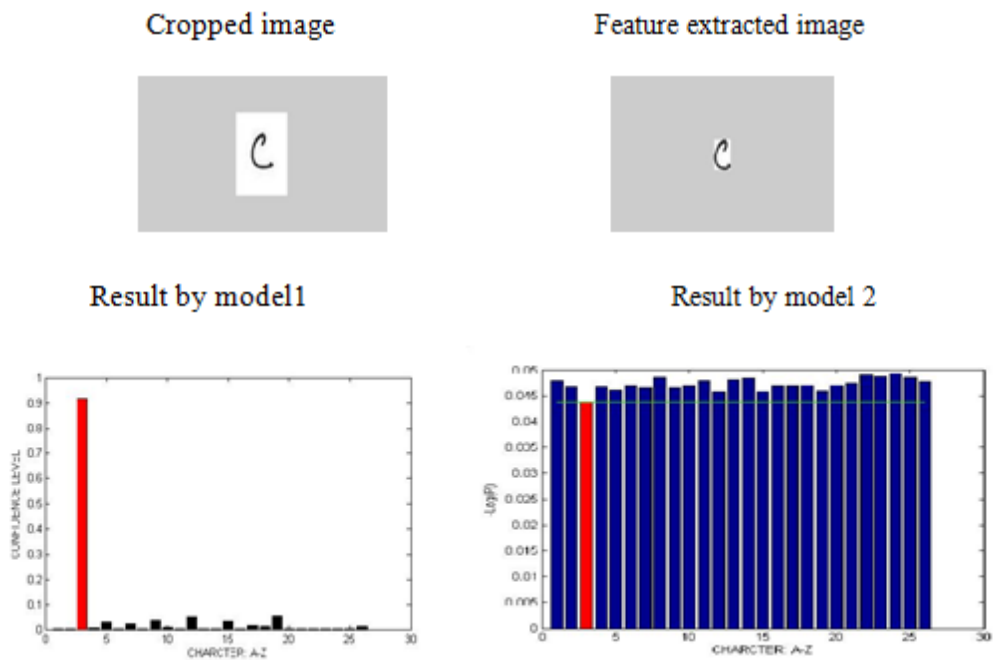

Result by model 3

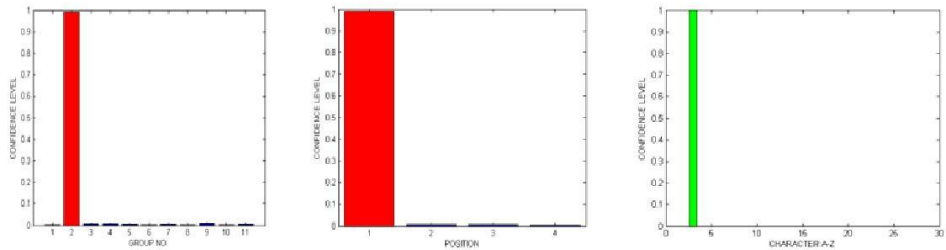

Figure-13.(a)performance in group recognition (b) performance in position recognition.
(c)character recognition

## CONCLUSION

Problem of manual approach of data entry has huge affect over quality of data in database and cost of processing. Automatic approach to recognize the data is considered as primary objective especially at front end where more manual approaches involved. To automate the solution, the concept of intelligence with image processing has applied in this paper. Various different modules using feedforward neural architecture have been developed to have diversity in applied intelligence. It has seen that presented direct and correlation based models work with nearly satisfactory and same performance where as hierarchical model has shown little less robustness. Proposed solution efficiency can be increased up to required level with help of controlled input environment where there is a fixed specified space available to write data and automatic cropping. It is anticipated that presented solution will appear as considerable solution for the problem of data cleaning at the front end.